\documentclass[11pt,twoside]{article}
\usepackage{asp2010}

\resetcounters
\usepackage{psfig}

\begin{document}

\title{The Decay of Stellar Dynamos and X-ray Activity}
\author{Nicholas J. Wright, Jeremy J. Drake \& Francesca Civano}
\affil{Harvard-Smithsonian Center for Astrophysics}

\begin{abstract}
Existing stellar X-ray surveys suggest major problems in our understanding of the evolution of stellar magnetic activity in solar and late-type stars, reaching conflicting conclusions over the rate of decay of X-ray activity and the spectral types responsible. We are confronting these discrepancies with a new model of the expected stellar X-ray luminosity distribution, combining a Galactic population synthesis model with current theories for rotational spin- down and the rotation -- activity relation for the stellar magnetic dynamo. Here we test our model using new observations of the stellar content of the {\it Chandra} COSMOS survey, for which 60 new stellar X-ray sources are identified from the thin disk and Galactic halo populations. Our model is in approximate agreement with the observed X-ray luminosity distribution and the distribution of spectral types responsible. However, slight differences in the form of the X-ray luminosity distribution exist that may hint at problems in our understanding of stellar X-ray emission.
\end{abstract}

\section{Introduction}

Solar and late-type stars emit X-rays through a magnetically confined plasma, or corona, at several million Kelvin. The corona is believed to be powered by the stellar dynamo, which itself is powered by differential rotation within the star \citep[e.g.][]{skum72,pall81,noye84}. The observed decrease in stellar X-ray luminosity of several orders of magnitude between the zero age main sequence \citep[e.g.][]{feig02,flac03,wrig10a} and solar age \citep[e.g.][]{pere00} has therefore been attributed to the rotational spin-down of the star, though a consistent picture has yet to emerge. X-ray observations of stellar clusters up to a billion years in age have been used to study the age -- rotation and rotation -- activity relationships \citep[e.g.][]{barn03,pizz03}. However, the dispersal of stellar clusters as they age means that very few stellar clusters older than a billion years exist, and many of these are faint and distant. Age estimates for individual main-sequence field stars are difficult to derive independently and therefore the evolution of both rotation and stellar activity for older stars is poorly understood.

Attempts to overcome this problem by using large samples of field stars have often resulted in conflicting conclusions. \citet{gude97} studied a sample of nearby solar-type stars aged $1-10$~Gyr and found that the X-ray luminosities decayed as $L_X \propto t^{-1.5}$, while \citet{mice02} could find no evidence for a clear decay law over a similar age range, and \citet{feig04} estimated a decay law of $L_X \propto t^{-2}$ from a sample of faint high Galactic latitude main-sequence stars. Other studies have attempted to understand the evolution of stellar X-ray emission by comparing predictions from models \citep[e.g. XCOUNT,][]{fava92} with X-ray surveys of field stars. However, these methods have also uncovered discrepancies such as an observed excess of yellow (G-K dwarf) stars compared to model predictions \citep{fava88}, and evidence for an excess of young main-sequence stars suggesting either a recently high star formation rate \citep{mice93} or uncertainties in the binary star fraction \citep{mice07}.

These discrepancies are no doubt partly influenced by the small but diverse range of stellar X-ray samples used. These include wide-field surveys from {\it Einstein} and {\it ROSAT} \citep[e.g.][]{schm04} that may be biased toward bright and nearby thin-disk stars as well as deep surveys with the {\it Chandra} and {\it XMM-Newton} observatories that include many distant or faint sources \cite[e.g.][]{feig04}. Between these two extremes deep and wide-field surveys such as the {\it Chandra} Cosmic Evolution Survey \citep[COSMOS,][]{elvi09} offer the balance necessary to produce large samples of stellar X-ray sources that are also deep enough to uncover intrinsically faint sources.

In this contribution we discuss a new model for predicting the X-ray emitting stellar population in any field of view of the Galaxy. We test this model against a new and well-constrained catalog of stellar X-ray sources detected in the {\it Chandra} COSMOS survey.

\section{XStar: a stellar X-ray emission model}

To further our understanding of the evolution of stellar dynamo activity we wish to compare the X-ray luminosity distribution of a well constrained sample with that predicted from current models of the X-ray evolution of late-type stars. Doing this for individual stars is difficult because of uncertainties in deriving accurate stellar ages. An alternative method that has been used in the past \citep[e.g.][]{fava92,mice93} is to compare the observed X-ray luminosity distribution with predictions from a model that takes into account the stellar populations along the line of sight and the spectral type dependent age -- rotation -- activity evolution of late-type stars. 

To do this we have a compiled a new X-ray population synthesis model, XStar, that combines the latest theories of Galactic structure, rotational evolution and stellar dynamo activity to make predictions about the X-ray emitting stellar population in any field of view. The ingredients of this model are as follows:
\begin{itemize}
\item A model of the spatial distribution and properties of stars in the Galaxy along a particular line of sight. For this we use the TRILEGAL population synthesis model \citep{gira05}, which can be calibrated using deep optical observations of the field in question.
\item A theory for the spectral-type dependent stellar spin-down as stars age. For this we use the framework outlined by \citet{barn03} and in particular the rotational isochrones of \citet{meib09}.
\item The X-ray activity -- rotation relationship quantified by \citet{pizz03}, using a new empirical determination of the convective turnover time based on rotation period measurements of field and cluster stars.
\end{itemize}

The product of this model is a probability distribution function of the X-ray emitting stars along the line-of-sight, including spectral types, distances, ages, rotation periods, and X-ray luminosities. This distribution can then be compared to the observed distribution of these quantities to allow the components making up the model to be tested. Where discrepancies exist between model and observations, different quantities can be compared to test different aspects of the model such as the Galactic star formation history (through the distribution of distances) or the binary star fraction (through the spectral type distribution). A fully calibrated model can then be used to make predictions of the X-ray emitting foreground population that may contaminate studies of star forming regions \citep[e.g.][]{wrig09a} or young clusters \citep{pill06}, or be used to assess the late-type stellar contribution to the extended X-ray emission from nearby galaxies \citep[e.g.][]{boro10}.

\section{Testing the model on the {\it Chandra} COSMOS field}

To test the XStar model we use the new catalog of stellar X-ray sources in the {\it Chandra} COSMOS survey presented by \citet{wrig10b}. The deep and abundant optical and near-IR photometry for the field allows the Galactic starcounts model to be accurately calibrated, while the completeness fractions estimated by \citet{pucc09} can be used to modulate the resulting X-ray distribution functions.

\subsection{New stellar X-ray sources in the {\it Chandra} COSMOS field}

\begin{figure}
\plotone[width=340pt]{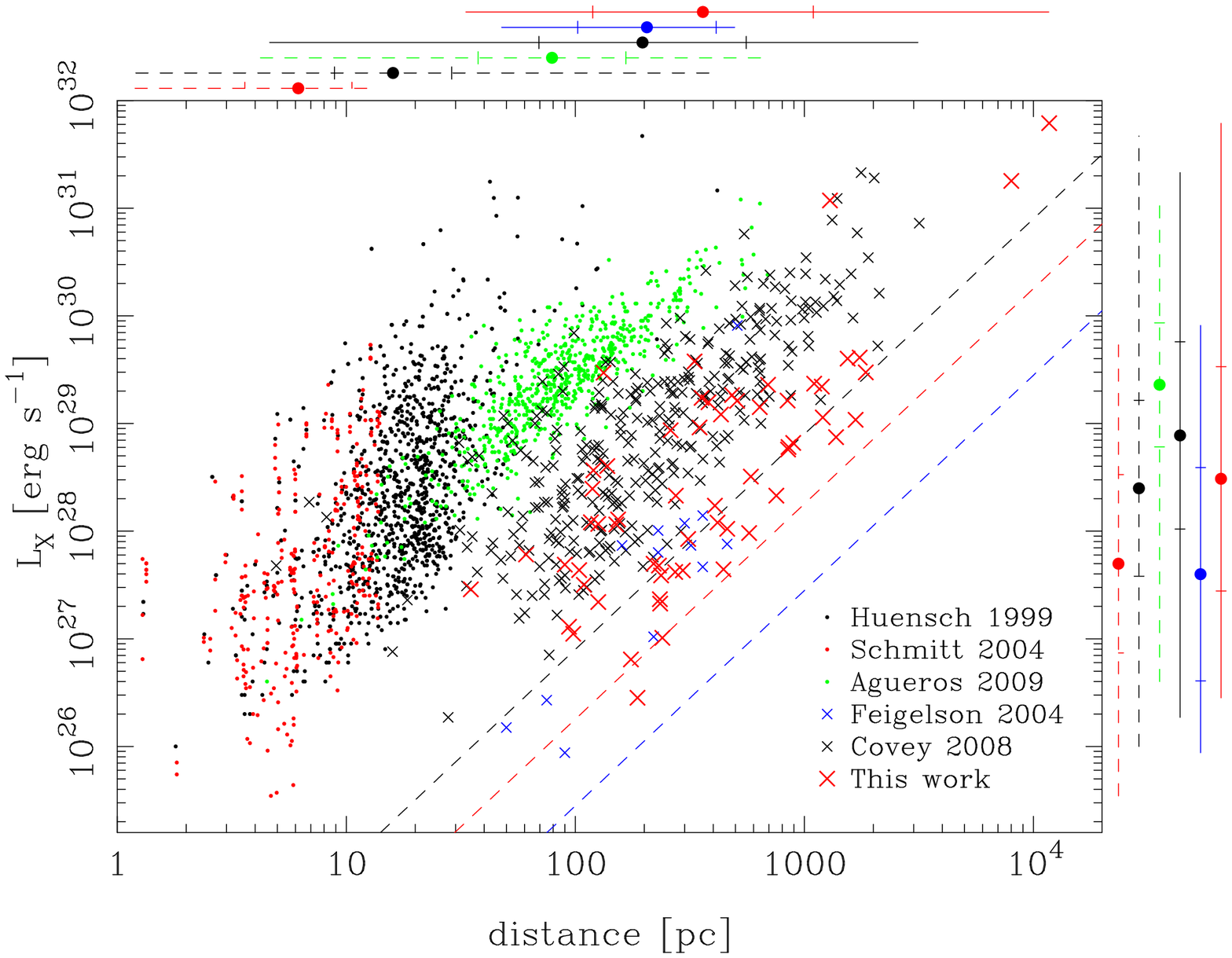}
\caption{$L_X$ vs distance for this and other samples \citep{huen99,schm04,feig04,ague09,cove08}. Dashed lines show estimated sensitivity limits \citep{feig05} for the three {\it Chandra} surveys (in the same color as their symbols). Also show are the range of distances (above) and $L_X$ (right) spanned by each sample (in the same color as their symbols), with the mean and standard deviations marked as dots and dashes respectively.}
\label{distances}
\end{figure}

The Cosmic Evolution Survey \citep[COSMOS,][]{scov07} is a deep and wide extragalactic survey designed to probe the medium redshift galaxy and active galactic nuclei populations. The field has been observed at nearly all wavelengths \citep[e.g.][]{capa07,scov07b} providing a large multi-wavelength catalog. The {\it Chandra} survey of the COSMOS field \citep{elvi09} has imaged an area of $\sim$0.9~deg$^2$ with a uniform exposure of $\sim$160~ks over the central 0.5~deg$^2$ and $\sim$80~ks over the outer region. A detailed source detection procedure \citep{pucc09} resulted in a catalog of 1761 sources with well defined completeness fractions as a function of both X-ray band and survey area.

Optical and near-IR identifications for 1750 of these sources were made by \citet{civa10}, including a number of stellar sources. \citet{wrig10b} used photometry and spectroscopy to confirm the stellar nature of 60 of these sources, obtaining spectral classifications, photometric distances and extracting X-ray properties such as plasma temperatures and luminosities. The sources uncovered span the full spectral sequence of coronal X-ray emitters, from late F-type to mid M-type, with distances ranging from 30~pc to 12~kpc, extending the known coverage of the $L_X$ -- distance plane to greater distances and higher X-ray luminosities than ever before (see Figure~\ref{distances}).

The majority of sources are more luminous than the contemporary Sun \citep{pere00} and the spectral-type dependence of the X-ray luminosities is in good agreement with previous studies \citep[e.g.][]{cove08}. The most distant of the sources are likely to be the most luminous members of the X-ray - emitting Galactic halo population, which appears to comprise both low-activity spectrally hard sources, that may be emitting through thermal bremsstrahlung, as well as a number of highly active sources in close binaries. Two sources are identified at distances of $\sim$10~kpc, which also have the highest X-ray luminosities in our sample, $L_X \sim 10^{31 - 32}$~erg~s$^{-1}$. We consider the possibility that these are either mis-identifications, confused sources or background fluctuations but conclude that they are more likely to be bona fide stellar X-ray sources. Their high X-ray luminosities, despite their considerable age, suggests that they are in close binaries that have been kept active through tidal interactions and the tapping of orbital angular momentum to sustain strong dynamo activity. See \citet{wrig10b} for a full analysis and discussion of this sample and its X-ray properties.

\subsection{Calibration of the Galactic starcounts model}

\begin{figure}
\plotone[width=350pt]{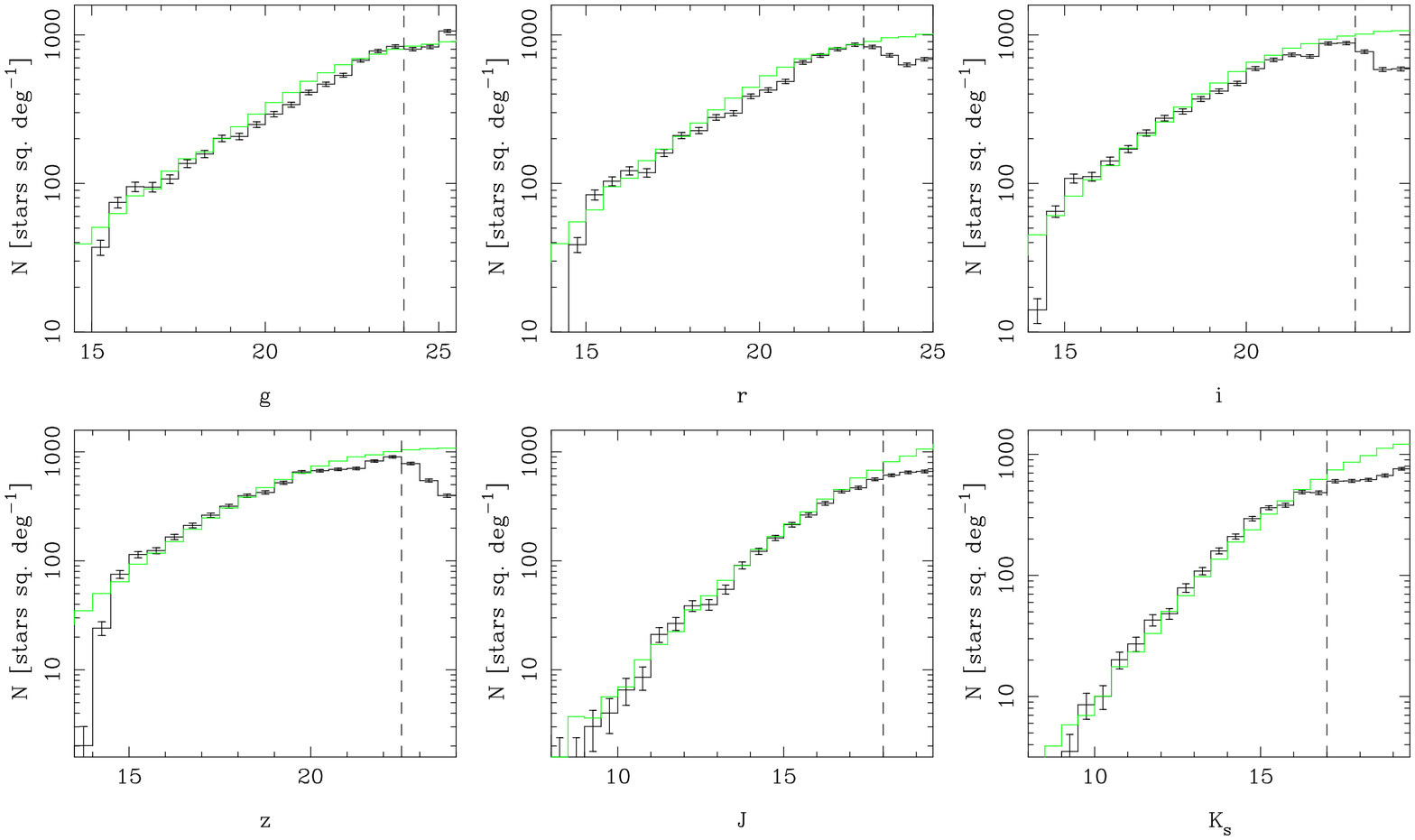}
\caption{Starcounts for the COSMOS field (black) and the best-fit TRILEGAL model (green) in the SDSS and 2MASS bands, with completeness limits and 1$\sigma$ Poisson uncertainties marked.}
\label{starcounts}
\end{figure}

A catalog of all stellar sources in the COSMOS field was compiled using data from the COSMOS optical catalogs \citep{capa07,mccr10}, the Sloan Digital Sky Survey \citep[SDSS,][]{york00} and the Two Micron All Sky Survey \citep[2MASS,][]{skru06}. All the photometry was converted onto a single photometric system based on the SDSS and 2MASS systems, and then filtered for galaxies using morphological and spectral energy distribution cuts based on the known locus of the main-sequence in the various two-color diagrams analyzed by \citet{cove07}. The product of these cuts was to reduce the original COSMOS catalog of 1.7 million sources initially to $\sim$500,000 point-like sources, and then to only $\sim$75,000 sources with stellar colors. The TRILEGAL Galactic star counts model \citep{gira05} was then used to fit a simple Galactic structure model to this sample (see Figure~\ref{starcounts}), altering a small number of parameters to reflect local differences in Galactic structure parameters that produce a better fit to the observations. The result of this population synthesis model is a list of stellar sources in the field of view with important properties such as their age, spectral type and binarity that can then used as the input in our stellar X-ray emission code.

\subsection{Results}

\begin{figure}
\plotone[width=300pt]{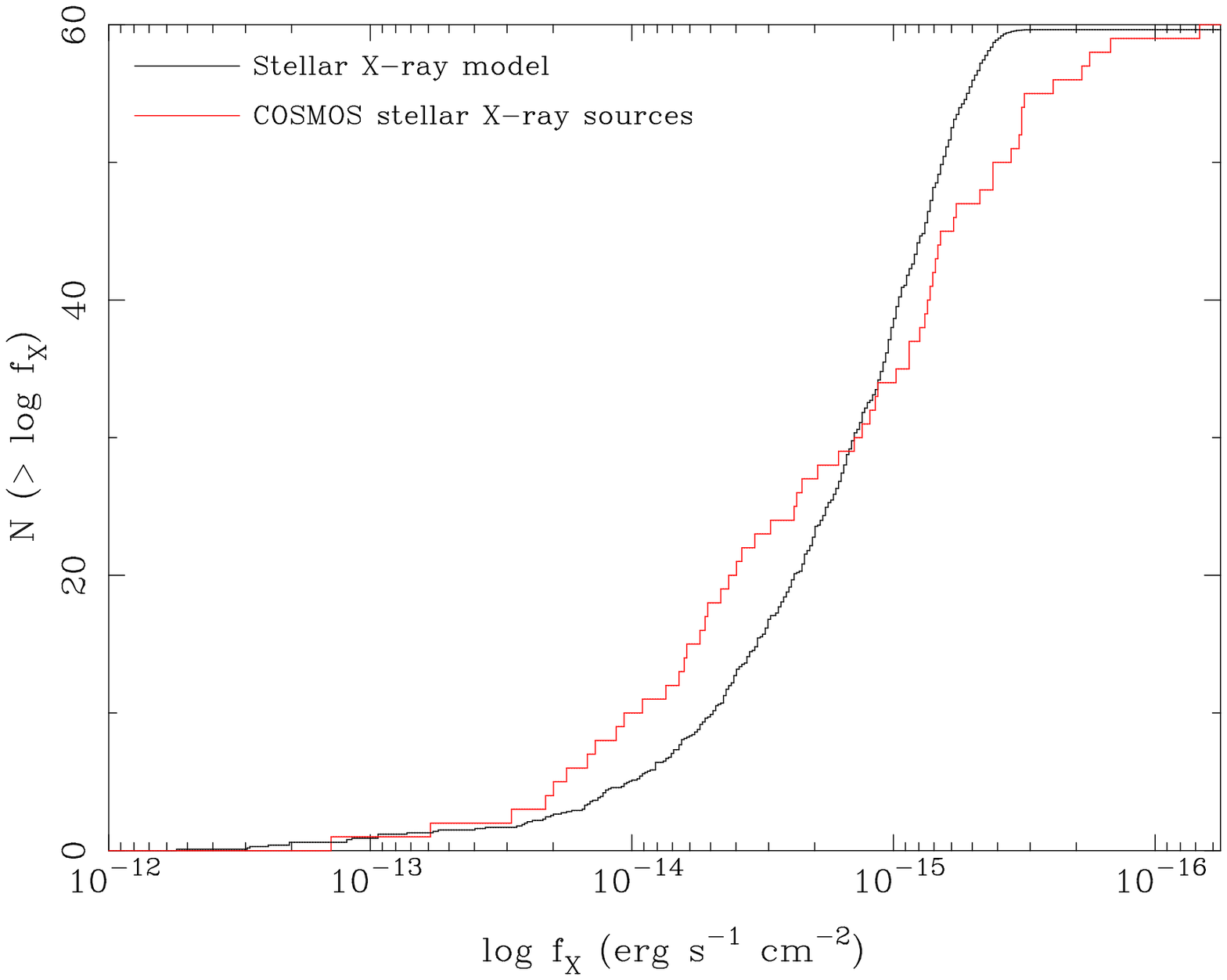}
\caption{Cumulative X-ray flux distribution for the COSMOS stellar sample and the results of our model (with the COSMOS completeness limits applied).}
\label{model}
\end{figure}

The results of our model are shown in Figure~\ref{model} compared to the observed cumulative X-ray flux distribution of sources in the COSMOS field. The model provides a reasonably good fit to the brightest sources, but under-predicts the number of sources with fluxes between $2 \times 10^{-14}$ and $2 \times 10^{-15}$~erg~s$^{-1}$~cm$^{-2}$. Fainter than this the model begins to over-predict the number of sources. However, below $f_X = 10^{-15}$~erg~s$^{-1}$~cm$^{-2}$ the COSMOS survey is incomplete \citep{pucc09}, and therefore differences between the model and observations could equally be attributed to uncertainties in the completeness curves for the COSMOS observations (which were determined based on the spectral energy distributions of galaxies and may therefore be inaccurate if used for stars).

Considering only the cumulative flux distribution above the completeness limit, the model under-predicts the number of sources at the 2$\sigma$ level. The spectral type distribution and distance distribution of the model and observations are in better agreement, suggesting that the discrepancy does not lie with some unresolved component of the model (e.g. a recently enhanced star formation rate), but with the age -- rotation -- activity relationships. Given that the age -- rotation relationship is better calibrated than the rotation -- activity relationship, the source of the discrepancy is more likely to be the latter. A steeper rotation -- activity relationship could partly resolve this discrepancy, such that in the simple approximation $L_X \propto v_{rot}^\beta$, $\beta > 2$ \citep[where $\beta = 2$ is the standard exponent used in the rotation -- activity relationship, e.g.][]{pizz03}. 

A similar steepening of this relationship was found by \citet{feig04} from their smaller sample of {\it Chandra} Deep Field North stars. They found a decay of X-ray activity with time of $L_X \propto t^{-2}$, which for slowly-rotating stars spinning down Skumanich-style ($v_{rot} \propto t^{-1/2}$) reduces to a rotation -- activity relation of $L_X \propto v_{rot}^{4}$. We find that this significantly over-predicts the number of observed stellar X-ray sources. An intermediate level of decay found by \citet{gude97} was $L_X \propto t^{-1.5}$, which is equivalent to $L_X \propto v_{rot}^{3}$ for slow rotators, and offers a better fit, but still over-predicts the number of sources. An intermediate value of $\beta \sim 2.5$ produces the best fit, though at the moment we can only rule out the standard case of $\beta = 2$ with a confidence of 2$\sigma$. Larger samples are clearly necessary to resolve this.

Whether this steepening of the rotation -- activity relation is universal or is an age-related effect is hard to determine. \citet{feig04} consider the possibility that magnetic field generation becomes more efficient as stars age or that the scaling of coronal densities behaves differently for older stars. However, the fact that this apparent steepening of the rotation -- activity relation has been observed in a number of samples that cover a range of ages \citep[our stars are more X-ray luminous and therefore likely to be younger than those found by][]{feig04} could indicate that this steepening is universal. \citet{mont01} argue that current models of two-layer interface dynamos \citep[e.g.][]{park93,char97} do not scale simply as $v_{rot}^2$, but are a more complex function of a number of parameters, not all of which can currently be measured or determined from models. Empirical estimates of this scaling can therefore better inform stellar dynamo models, ultimately helping us better understand our own Sun.

\section{Conclusions}

To improve out understanding of the decay of stellar dynamo activity we have compiled a new model of stellar X-ray activity, XStar, which incorporates a population synthesis Galactic model, the theories of spectral-type dependent rotational spin-down and the rotation -- activity relationship. The results of the model can be compared to samples of stellar X-ray sources from a range of observatories. Here we test the model using new observations of 60 stellar X-ray sources identified in the {\it Chandra} COSMOS field. A comparison of the distribution of X-ray fluxes, spectral types, and distances for the observed and modeled populations shows a generally good agreement, but slightly under-predicts the number of observed sources. Differences between the two X-ray flux distributions suggests that the our current formulation of the rotation -- activity relationship may not be correct, and a steeper relationship may better describe the observations.

\acknowledgements 

NJW was funded by {\it Chandra} grant AR9-0003X and JJD was supported by NASA contract NAS8-39073 to the {\it Chandra} X-ray Center. FC is supported by NASA {\it Chandra} grant number G07-8136A to the {\it Chandra} COSMOS Survey. We thank the COSMOS team for their work on this survey and assistance with this research. We are also grateful to the staff at the Fred Lawrence Whipple Observatory for spectroscopy used in this work.

\bibliography{/Users/nwright/Documents/Work/tex_papers/bibliography.bib}

\end{document}